\documentclass{aa}  

\usepackage{graphicx}
\usepackage{subfigure}
\usepackage{subfloat}
\usepackage{txfonts}
\begin{document} 
   \title{Grain growth in the envelopes and disks of Class I protostars}


   \author{A. Miotello
          \inst{1}\fnmsep\inst{2}
          \and
          L. Testi \inst{1}\fnmsep\inst{3}\fnmsep\inst{4}
          \and
          G. Lodato\inst{2}
          \and
          L. Ricci\inst{5}
          \and
          G. Rosotti\inst{4}\fnmsep\inst{6}
	\and
	K. Brooks\inst{7}
	\and
	A. Maury\inst{8}
           \and
          A. Natta\inst{3}\fnmsep\inst{9}
          }

   \institute{European Southern Observatory (ESO), 
   	   Karl-Schwarzschild-Strasse 2, D-85748 Garching, Germany              
         \and
             Universita' degli Studi di Milano, Dipartimento di Fisica,
             Via Celoria 16, I-20133 Milano, Italy
         \and
            INAF-Osservatorio Astrofidico di Arcetri, Largo E. Fermi 5, I-50125 Firenze, Italy             
         \and
            Excellence Cluster Universe, Boltzmannstr. 2, D-85748 Garching, Germany             
          \and
             Department of Astronomy, California Institute of Technology,
             MC 249-17, Pasadena, CA 91125, USA
          \and
             Max-Planck-institute f{\"u}r extraterrestrische Physik, 
             Giessenbachstra{\ss}e, D-85748 Garching, Germany
	\and
	   Australia Telescope National Facility, P.O. Box 76, Epping NSW 1710, Australia 
	\and
	   Harvard-Smithsonian Center for Astrophysics, 60 Garden Street, Cambridge, MA 02138, USA
          \and
              School of Cosmic Physics, Dublin Institute for Advanced Studies, 31 Fitzwilliam Place, Dublin 2, Ireland}

  \abstract{
We present new 3 mm ATCA data of two Class I Young Stellar Objects in the Ophiucus star forming region: 
Elias29 and WL12. For our analysis we compare them with archival 1.1~mm
SMA data. In the $(u,v)$ plane the two sources present a similar behavior: 
a nearly constant non-zero emission at long baselines, which suggests the 
presence of an unresolved component and an increase of the fluxes at  
short baselines, related to the presence of an extended envelope. 
Our data analysis leads to unusually low values of the spectral index 
$\alpha_{\rm 1.1-3mm}$, which may indicate that mm-sized dust grains have 
already formed both in the envelopes and in the disk-like structures at such early stages. 
To explore the possible scenarios for the interpretation of the sources we perform a radiative transfer modeling using a Monte Carlo code, in order to take into account possible deviations from the Rayleigh-Jeans and 
optically thin regimes. 
Comparison between the model outputs 
and the observations indicates that dust grains may form aggregates
up to millimeter size already in the inner regions of the envelopes 
of Class I YSOs. Moreover, we conclude that the embedded disk-like structures in our 
two Class Is are probably very compact, in particular in the case of WL12, 
with outer radii down to tens of AU. 
}


   \maketitle
%

\section{Introduction}
Stars are formed out of the gas present in dense molecular clouds. In 
particular small scale cores (with sizes of the order of 0.1 pc and 
estimated lifetime $\sim$ 10$^6$ yrs) within molecular clouds have been 
identified as the first stage of star formation \citep[see review by][]{Bergin07}. The new-born protostars 
are initially embedded in rotating and collapsing envelopes made of cold 
gas and dust, which form in those cores. The material is gradually accreting from the 
envelopes onto the central objects through  circumstellar disks, which 
form due to the conservation of angular momentum. Disks are thought to be the birthplace of planets. For the formation of rocky planets like our Earth, 
and probably also for rocky cores of giant planets, the dust component 
has to grow from the submicron-sized interstellar medium (ISM) grains 
\citep{Mathis} up to thousands of kilometers-sized bodies. During the 
first stages of the grain growth process gas and dust interact, 
leading the solid particles to collide and finally to aggregate 
\citep[see the extensive reviews in][]{Beckwith00,Natta07,Testi13}.

Young stellar objects (YSOs) are classified according to the slope of their spectral energy distributions (SEDs), usually in the range of wavelengths from 2.2 $\mu$m to 20 $\mu$m. Initially YSOs were studied in the infrared (IR) \citep{Lada87}, and were classified as  Class I, Class II and Class III YSOs, depending on their near infrared (NIR) slope. Subsequently, Class 0 YSOs were also defined observationally by \cite{Andre93}.

This observational classification is usually interpreted as an 
evolutionary sequence.
In particular Class 0 YSOs are considered to be the first stage, when 
the prostostar is still completely embedded in its envelope that 
absorbs and reemits the radiation from the central star and disk (and $M_{\rm env} \gg M_{\rm proto-\star}$). They evolve towards the Class I 
phase, where the envelope is partially dissipated ($M_{\rm env} \lesssim M_{\rm proto-\star}$) and the emission 
from the primordial disk can be better separated from that of the envelope. In the Class II stage the envelope has disappeared, the central 
pre-main-sequence star (PMS) becomes optically visible, and the disk 
is optically thick in the IR and optically thin in the mm. Then the 
source enters in the Class III phase, where the more evolved PMS may 
be surrounded by a tenuous, optically thin disk and planets or 
a ``debris disk'' may be present.

An interesting question is when and where dust grains begin to grow 
during this evolutionary sequence. Observational results indicate that 
nearly all the disks in the Class II phase contain dust grains with 
sizes of at least 1 mm in their outer regions 
\citep[see e.g.][and references therein]{Ricci}, 
independently on mass and environment
\cite[note however the interesting case of the Chamaeleon~I region by][]{Ubach2012}. 
This shows that the formation of these large grains has to be very 
rapid, occurring before a YSO enters in the Class II stage, in the 
early phases of disk formation or possibly even in the 
collapsing envelopes.

\cite{Pagani11} present a preliminary evidence of grain growth up to $\sim$1 $\mu$m already within dense cores, through the detection of a IR ``core-shine": strong scattering of background radiation due to micron-sized grains within the core. Similar levels of grain growth have also been derived from the study of the extinction law in Perseus dense cores \citep{Foster13}. Moreover according to the studies by \cite{Jorgensen07}, \cite{Kwon10} and \cite{Chiang12} there are possible indications from millimeter observations that even larger grains are present in the central regions of the evelopes surrounding Class 0 protostars. There is however a caveat in studying Class 0s, i.e. the emission may come from regions that are optically thick at sub-millimeter wavelengths.\\
Class I YSOs represent the transition phase between cores and disks, where both the envelope and the disk are detectable and the two different components can be separated more easily than in Class 0 YSOs. They are a well suited to study the first stages of grain growth and to constrain the initial condition for the evolution of protoplanetary disks.

The  radiative emission by dust grains depends on the typical grain size.
Dust continuum emission in the millimeter allows to place constraints 
on dust properties of disks and envelopes in nearby star forming regions 
(SFRs). As opposite to optical and infrared studies that are mostly 
sensitive to growth up to micron-size particles, millimeter wavelengths 
allow to test the presence of mm/cm-sized grains. 
At these long wavelengths the emission of the disk is mostly optically 
thin and, as a consequence, the spectral index $\alpha_{mm}$ of the SED 
($F_\nu \approx \nu^{\alpha_{mm}}$) carries the information on the 
spectral index $\beta_{\rm mm}$ of the dust opacity coefficient 
($\kappa_\nu \approx \nu^{\beta_{\rm mm}}$; see 
\citealt{BS91,Testi01,Natta07,Draine}). In particular, for the 
approximate case of optically thin emission in the Rayleigh-Jeans 
regime, $\alpha_{\rm mm}$ = $\beta_{\rm mm}$+2. ISM-like particles are 
usually characterized by $\beta_{mm}\sim$ 1.7 (and $\alpha_{mm}\sim$ 3.5), 
while larger mm/cm-sized grains have lower values of $\beta_{\rm mm}$ 
and $\alpha_{\rm mm}$ \citep{Draine}. This 
relation between $\alpha_{\rm mm}$ and $\beta_{\rm mm}$ holds 
independently on the geometry of the system, i.e. also in the 
envelope  $F_{\nu}\sim \nu^{2+\beta_{mm}}$.
Therefore, multi-wavelength observations of YSOs in the (sub-)millimeter 
can reveal the presence of mm/cm-sized pebbles in disks and in envelopes 
\citep{Testi13}.

In this paper we present new mm and cm observations of Elias29 and WL12, two Class I YSOs in the $\rho$ Ophiuchi SFR, with the aim of constraining the grain size distribution on the disk and envelope of each object separately. These objects have been studied previously with the Submillimeter array (SMA) at 1.1 mm by \cite{Jorgensen}.
In Sect. \ref{observations} we present our new observations of both Elias29 and WL12 with the Australian Telescope Compact Array (ATCA) at 3 mm, 3 and 6 cm. Then, in Sects. \ref{results} and \ref{analysis} we present the results of the data reduction process and a first order analysis of the dataset. A more accurate analysis is presented in Sect. \ref{modeling} and discussed in Sect. \ref{discussion}. Finally we summarize our results in Sect. \ref{Summary}.


\section{Observations}
\label{observations}
\subsection{Sources selection}
We selected two sources located in the $\rho$ Ophiuchi SFR, taken to be at a distance D = 125 $\pm$ 25 pc \citep{deGeus,Lombardi08}: Elias29 and WL12. Their infrared colors, their bolometric luminosities and temperatures were obtained with Spitzer photometry from the "Cores to Disk" legacy program \citep{Evans09}. The two sources have similar bolometric temperatures, $T_{\rm bol}$ = 391 K for Elias29 and $T_{\rm bol}$ = 440 K for WL12, while their bolometric luminosities are estimated to be $L_{\rm bol}$ = 13.6 $L_{\odot}$ and $L_{\rm bol}$ = 2.6 $L_{\odot}$, respectively.\\
The two sources have been classified as Class I from the slope of their SED in the near- and mid-infrared \citep{Evans09}.\\
They have been observed also at 1.1 mm with SMA within the PROSAC program \citep{Jorgensen}, and in particular Elias 29 has been studied by \cite{Lommen}, who gave an estimate of the star, disk and envelope masses (Table \ref{par_lit}).\\
\subsection{ATCA  observations}
The observations were performed with the Australian Telescope Compact Array (ATCA) using the CABB wide band correlator between July 2011 and September 2011 in three variations of its hybrid configurations: H75, H168 and H214. Two 2 GHz wide band centered at 93.0 and 95.0 GHz were used for the 3 mm observations, while for 3 and 6 cm we used 2 GHz wide band  centered at 9.0 and 5.0 GHz, respectively.  The projected baselines provided were 7 to 79 k$\lambda$ for Elias29, while they were 7 to 98 k$\lambda$ for WL12. Calibration was done using the MIRIAD package \citep{Sault}. The sources 1253-055 and 1622-297 served as band-pass and phase calibrators and the absolute fluxes were calibrated on Uranus.
\subsection{SMA  observations}
In our analysis we made also use of archival 1.1 mm data (frequencies: 267 and 277 GHz),  obtained with SMA within the PROSAC program \citep{Jorgensen}.
The observations were performed between May 2006 and July 2007 in two variations of SMA compact configuration. The projected baselines were 5.9 to 64 k$\lambda$ for Elias29 and  9 to 104 k$\lambda$ for  WL12.
\begin{table}[ht]
\caption{Parameters from literature.}
\centering
\begin{tabular}{lcc}
\hline \hline
\textbf{Source} & \textbf{Elias29} & \textbf{WL12} \\
\hline
RA[J2000] & 16 27 09.4 & 16 26 44.2 \\
Dec[J2000] & -24 37 20.0 & -24 34 48.7 \\
$L_{\rm bol}[L_{\odot}]$ & 13.6 $^*$& 2.6$^*$ \\
$T_{\rm bol}[\rm K]$ & 391$^*$ & 440$^*$ \\
$M_{\rm s}[M_{\odot}]$ & 2.5 $\pm$0.6$^{**}$ & - \\
$M_{\rm disk}[M_{\odot}]$ & $\leq$ 0.007$^{**}$& 0.011$^*$\\
$M_{\rm env}[M_{\odot}]$ & $\leq$ 0.058$^{**}$ & 0.054$^*$ \\
\hline
\label{par_lit}
\end{tabular}
\tablebib{
$^*$\cite{Jorgensen};$^{**}$\cite{Lommen}}
\end{table}

\section{Results}
\label{results}

\subsection{Results at mm wavelengths}

\begin{figure*}
   \resizebox{\hsize}{!}
             {\includegraphics[width=1.\textwidth]{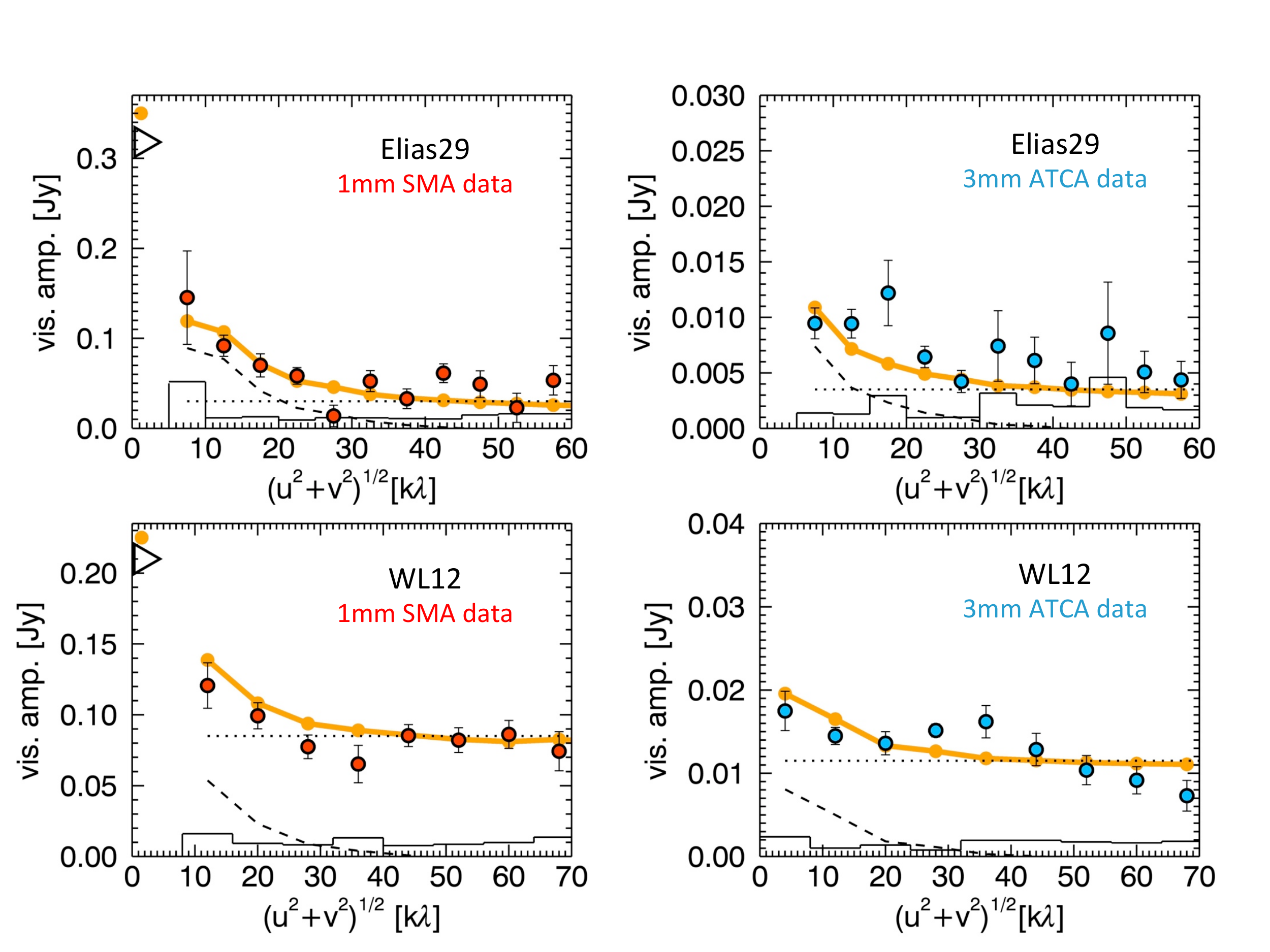}}
      \caption{Visibility amplitudes expressed in Jansky against the \emph{u-v} distances in k$\lambda$.The data points give the amplitudes per bin. The data are binned in annuli according to \emph{u-v} distance. In the top panels we present the amplitudes of Elias29: on the left hand side, the archival 1.1 mm SMA dataset \citep{Jorgensen}, while on the right the new 3 mm ATCA dataset. In the bottom panels we plot the same for WL12. The histograms give the expected amplitudes for zero signal. The open triangles at zero \emph{u-v} distance give the zero spacing 1.1 mm fluxes, interpolated between 850 $\mu$m and 1.25 mm single dish fluxes \citep{Lommen,Jorgensen}. The orange lines are the best fits of the observed emission, obtained using a combination of a two-layer disk model \citep[][]{Chiang, Dullemond} surrounded by a power-law density envelope. We used the Monte Carlo radiative transport model RADMC-3D \citep{RADMC3D} to compute the emission from the envelope. The separate contributions given by the disk and the envelope are plotted with the dotted and dashed lines respectively}
         \label{vis_mod}
   \end{figure*}

The visibility amplitudes at 3 mm  are plotted in the right panels of Fig. \ref{vis_mod} as a function of projected baselines, 7 to 60 k$\lambda$ for Elias29 and 7 to 70 k$\lambda$ for WL12. We report also the visibility amplitudes obtained with the archival 1.1 mm SMA data (left panels), whose baselines range from 5.9 to 60 k$\lambda$ for Elias29 and from 9 k to 70 k$\lambda$ for WL12.
The observations at 3 mm have been taken using a range of \emph{u-v} distances chosen to overlap with those of the SMA dataset. The smallest physical scales that we are able to resolve at both 1.1 and 3 mm are around 400 AU, while the short baselines correspond to scales larger than a few thousands of AU. 

In the case of Elias29 the emission at 1.1 mm rises at the shortest baselines, indicating the presence of an envelope, as already concluded by \cite{Lommen}. Indeed the \emph{u-v} distances at which the fluxes start to increase correspond to physical scales too large to be related with a resolved disk. The increasing trend is not as evident in the 3 mm plot, because of the lack of data for projected baselines shorter than 7 k$\lambda$. At larger \emph{u-v} distances on the other hand the emission is approximately constant in both data sets, showing the presence of an unresolved source, i.e. a disk-like structure. A similar behavior is found for WL12, even if the rise is less steep at the short baselines.

\begin{table}[ht]
\caption{Source fluxes at different wavelengths.}
\centering
\begin{tabular}{lcc}
\hline \hline
\textbf{Source} & \textbf{Elias29} & \textbf{WL12} \\
\hline
F$_{\text{3mm}}$[mJy] & 10.36$\pm$0.18 & 17.48$\pm$0.22 \\
F$_{\text{3cm}}$[mJy] & < 0.54 & < 0.55 \\
F$_{\text{6cm}}$[mJy] & < 0.26 & < 0.37 \\
\hline
\label{par_res}
\end{tabular}
\end{table}

\subsection{Estimated ionized gas contribution}
\label{cm}
\begin{figure}
   \resizebox{\hsize}{!}
           {\includegraphics[width=0.5\textwidth]{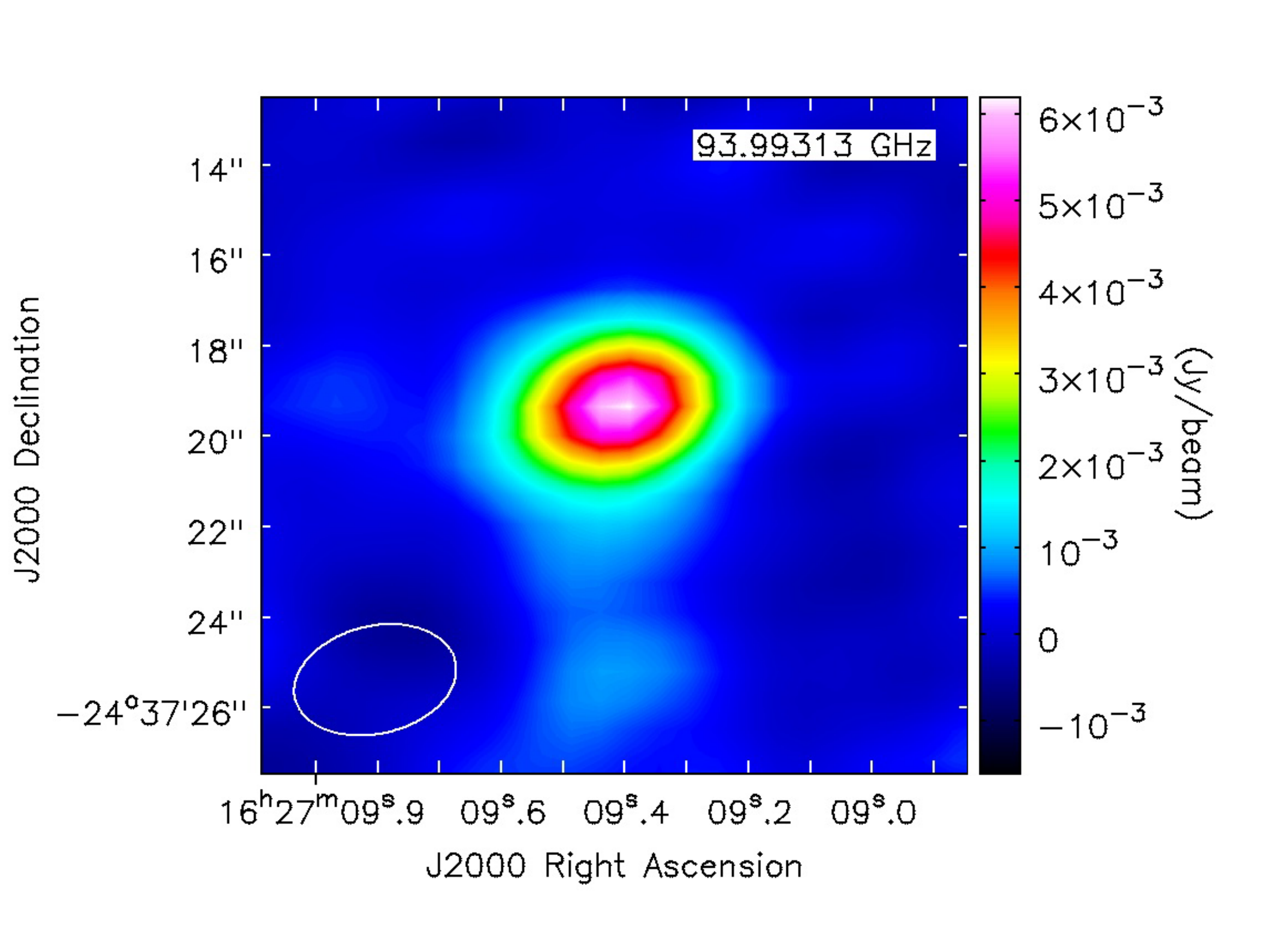}}
      \caption[Elias29 Map (3 mm).]{Elias29 map: detection of the source at 3 mm. The total flux of the source at 3 mm is 10.36 mJy, with a 3$\sigma$ rms of 0.18 mJy.}
         \label{Elias29map3mm}
 \end{figure}

\begin{figure}
   \resizebox{\hsize}{!}
              {\includegraphics[width=0.5\textwidth]{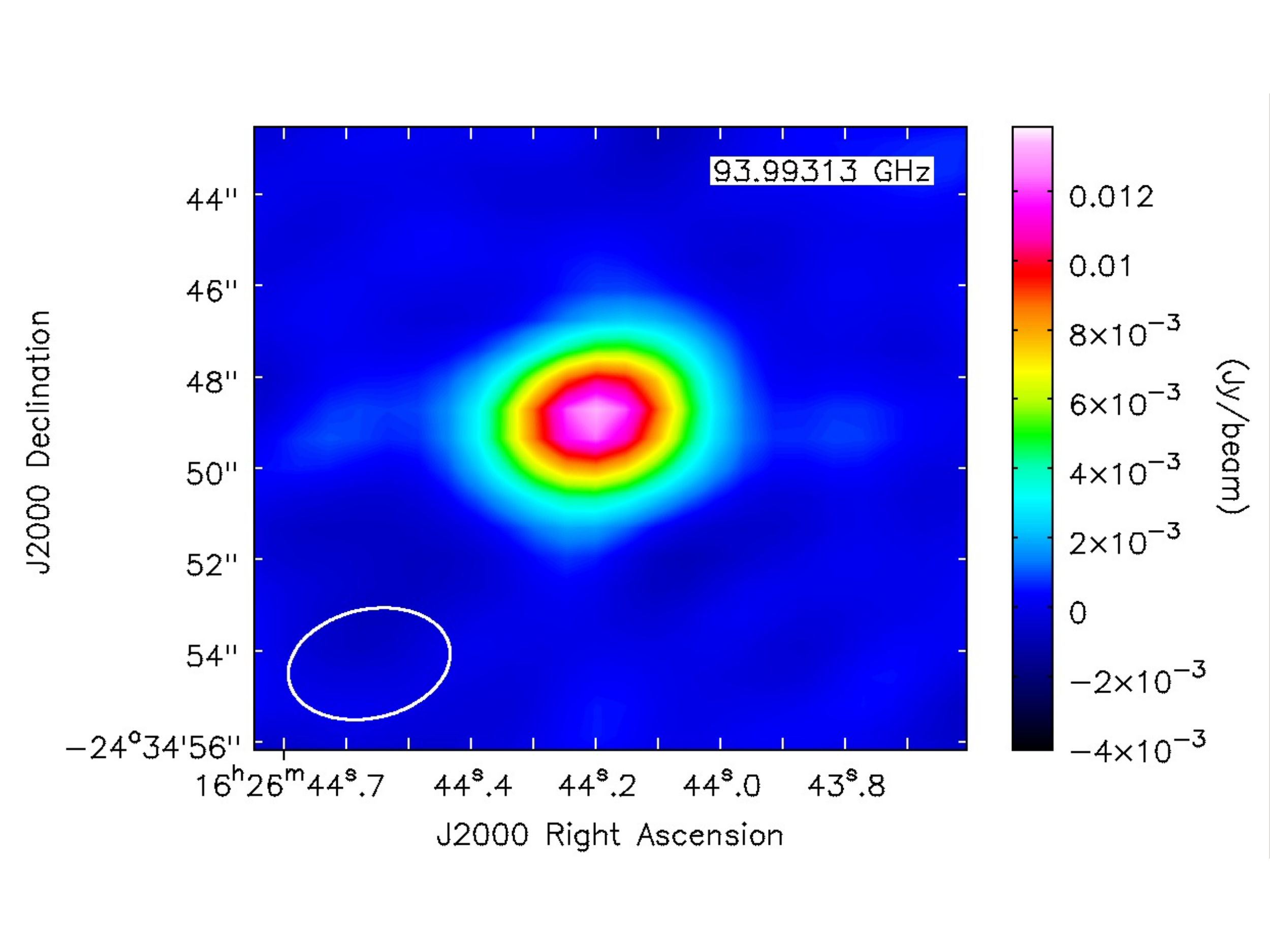}}
    \caption[WL12 Map (3 mm).]{WL12 map: detection of the source at 3 mm. The total flux of the source at 3 mm is 17.48 mJy, with a 3$\sigma$ rms of 0.22 mJy.}
         \label{WL12map3mm}
 \end{figure}

We obtained a clear detection of the two sources at 3 mm (Figures \ref{Elias29map3mm} and \ref{WL12map3mm}). On the other hand, we did not detect any of the two sources at 3 cm and at 6 cm. \cite{Dzib2013} have recently published a deep cm continuum survey of the $\rho$-Ophiuchi star forming region using the VLA. They 
detected Elias~29 ( J162709.41-243719.0) but not WL12. Their upper limits at 6 and 4~cm are a factor of $\sim$10 lower than ours. The measured cm-wave fluxes of Elias~29 are $\sim$0.25~mJy at 6~cm and $\sim$0.36~mJy at 4~cm, with a spectral index of $\sim$ 0.7. The sources is reported to be variable at the 30\%\ level. The extrapolated contribution of the gas emission at 3~mm would be $\sim$2~mJy, about 20\%\ of the total flux. We decided not to subtract this possible contribution from the 3 mm fluxes we measured, reported in the next paragraph, because of the source variability and because our observations and those of \cite{Dzib2013} are not simultaneous. The VLA observations were obtained in the period February-May 2011, while our own ATCA data were acquired in June, August and September 2011. The time difference between the various epochs is not large, but \cite{Dzib2013} report variability on a monthly timescale.  


To provide a worst case scenario for the contamination of the gas emission at 3 mm we use our simultaneous upper limits at 3 and 6 cm. The most efficient way to produce ionized gas which can contaminate the dust thermal emission at 3 mm is given by the presence of a stellar wind (i.e. mass ejection from the star). In this case the radio spectrum is found to vary as $F_{\nu} \propto \nu^{0.6}$ \citep{Panagia75}, while, for example, the free-free emission coming from an optically thin HII region would give $F_{\nu} \propto \nu^{-0.1}$ \citep{Mezger}. 
Using the 3 $\sigma$ limits at 3 and 6 cm (Table \ref{par_res}) we then compute the maximum possible contribution from an ionized wind at 3 mm.
The extrapolated contributions from ionized gas at 3 mm are $F^{\text{wind}}_{\text{3mm}}<$ 3.59 mJy for Elias29 and $F^{\text{wind}}_{\text{3mm}}<$ 3.66 mJy for WL12, less than 36$\%$ and 20$\%$ of the observed total fluxes at 3 mm respectively.

In Sect. \ref{analysis} we use our upper limits to estimate the impact of the free-free emission on the 
measurement of the spectral index of the dust emission, this choice corresponds to a larger maximum estimated contamination and is thus a conservative estimate. 


\section{Analysis}
\label{analysis}
As a first order analysis, we assume our sources to be composed by two elements, the envelope and the disk, whose emission are expected to be easily separable \citep[see e.g.][]{Jorgensen}. Since our observations do not allow us to resolve the disk (60 k$\lambda$ are related to disks diameters of the order of almost 400 AU) we expect the disk amplitude to be constant (unresolved source), without changing appreciably with baseline length. On the other hand we expect the envelope to be resolved and to dominate the emission on the largest detected scale. Therefore its emission should be high at the small baselines and approaching zero at the larger ones, where the disk contribution is expected to dominate. 
An analysis of the SMA data similar to this suggested the presence of extended envelopes and compact disks around these two sources \citep[see ][]{Lommen,Jorgensen}, as it also has for other sources.

For Elias29 in particular the value of the amplitudes is constant within the errorbars for baselines larger than 25 k$\lambda$ both at 1.1 and 3 mm. We assume this to be the emission coming from the embedded disk-like structure whose average values at the two wavelengths are: $F_{\rm disk}^{\rm 1mm}= 35.0 \pm 4.8 \,\text{mJy}$ and $F_{\rm disk}^{\rm 3mm}=5.0 \pm 0.7 \,\text{mJy}$. These values were obtained using a weighted average using the weights associated to the single observed visibilities. For WL12 we make the same assumption, relating the disk contribution to the amplitudes at baselines larger than 40 k$\lambda$. The disk average fluxes at the two wavelengths are: $F_{\rm disk}^{\rm 1mm}= 74.9\pm 4.6 \,\text{mJy}$ and $F_{\rm disk}^{\rm 3mm}=14.8\pm 0.7  \,\text{mJy}$. 

Under the optically thin assumption and the Rayleigh-Jeans approximation, given a dual-frequency observation where the fluxes at the frequencies $\nu_1$ and $\nu_2$ are $F_1$ and $F_2$, $\beta$ can be easily obtained as:
\begin{equation}
\beta = \frac{\ln F_1 - \ln F_2}{\ln \nu_1 - \ln \nu_2} - 2\, .
\label{beta_thin}
\end{equation}
Using the fluxes in the same \emph{u-v} ranges at 1 and 3 mm, we obtain therefore the average value of $\beta$ in the disk, which is $\beta_{\rm disk}= -0.16 \pm 0.19$ for Elias29 and $\beta_{\rm disk}= -0.47 \pm 0.07$ for WL12. These estimates assume that all the emission at 3 mm is due to thermal dust emission. We can use the upper limits on the cm-wave emission (see sect. \ref{cm}) to estimate the maximum possible contribution of the gas emission at 3mm. If we assume that there is a compact gas component emission at a level equivalent to the 3$\sigma$ upper limit derived from our cm observations (Table \ref{par_res}), then the value of $\beta$ in the disk becomes higher for both sources: $\beta_{\rm disk}^{\rm extr}= 1.1 \pm 0.5$ for Elias29 and $\beta_{\rm disk}^{\rm extr}= -0.21 \pm 0.08$ for WL12.

Second, we find an excess of emission at the very short baselines (i.e. 
from 6.9 k$\lambda$ up to 10 k$\lambda$ for Elias29 and from 6.8 k$\lambda$ up to 15 k$\lambda$ for 
WL12), when subtracting the disk visibilities from the total visibilities. The physical scales related to these \emph{u-v} distances range from around 2000 AU to 4000 AU.
The excess values at these short baselines are: $F_{\rm ex}^{\rm 1mm}=63\pm 15\,\text{mJy}$ and $F_{\rm ex}^{\rm 3mm}=4.1\pm 0.9\,\text{mJy}$ for Elias29, $F_{\rm ex}^{\rm 1mm}=28\pm 15\,\text{mJy}$ and $F_{\rm ex}^{\rm 3mm}=1.4\pm 0.6\,\text{mJy}$ for WL12. The excesses, even if they are not very large in the ATCA datasets, are detected especially in Elias29 and can be linked with the presence of an envelope in both sources. This hypothesis is also supported by the single-dish fluxes extrapolated at 1.1 mm combining 1.25 mm fluxes \citep{Andre94} and 850 $\mu$m fluxes \citep{Johnstone04,Ridge06} as done by \cite{Lommen}. They are shown by the open triangles in Fig. \ref{vis_mod}, and are substantially larger than the interferometer fluxes. They indicate that extended emission is present around the compact component source picked up in both the sources by the long baselines. This extended emission is due to the circumstellar envelope, as discussed by \cite{Lommen} and \cite{Jorgensen}.
Considering only the interferometer fluxes at these very short baselines, using the same \emph{u-v} distance ranges at 1.1 and 3 mm, it is possible to obtain an average value of $\beta$ given by the excess fluxes, accordingly with equation (\ref{beta_thin}), which are: $\beta_{\rm env}= 0.6 \pm 0.3$ for Elias29 and $\beta_{\rm env}= 0.8 \pm 0.7$ for WL12.
The errors on $\beta$ are estimated following the procedure shown in Appendix \ref{AppendixB}. The uncertainties on the values of $\beta_{\rm env}$ are large, since the visibilities are few and the S/N is accordingly very low.

Our inferred values of $\beta$ for both the disks and the envelopes are surprisingly low if compared to typical values found in the ISM, as reported in the introduction. To investigate the possible explanations, proper radiative transfer modeling of both structures is needed, taking into account possible deviations from the Rayleigh-Jeans and optically thin regimes, which can affect the determined $\beta$ values.

\begin{table}[ht]
\caption{Estimated stellar parameters.}
\centering
\begin{tabular}{lcc}
\hline \hline
\textbf{Source} & \textbf{Elias29} & \textbf{WL12} \\
\hline
$T_{\rm eff}[\rm K]$ & 4786 & 3980  \\
$R_{\rm eff}[R_{\odot}]$ & 5.9 & 3.5 \\
$M_{\rm eff}[M_{\odot}]$ & 3 & 0.6 \\
\hline
\label{par_birth}
\end{tabular}
\end{table}


\section{Modeling}
\label{modeling}

The modeling approach used in this work consists in separating the disk+protostar system from the envelope, as extensively done in previous studies \citep[e.g][]{Butner94,Hogerheijde99,Looney03,Jorgensen04,Jorgensen}. The disk+protostar system is modeled adopting the two-layer model by \cite{Dullemond}, whose output spectrum is taken as central source of heating in the envelope modeling. The envelope is modeled as done by \cite{Crapsi}, using the Monte Carlo Radiative Transfer code RADMC-3D \citep{RADMC3D}. Below we discuss in more details the disk and the envelope models separately.

\subsection{Disk models}
We adopt the two-layer model of a flared disk heated by the central protostellar radiation by \cite{Chiang}, as developed by \cite{Dullemond}. These models have been used by e.g. \cite{Testi03}, \cite{Natta04} and \cite{Ricci10} and we refer to these papers for a detailed description of the models.

First, in order to characterize a model, we need to specify the properties of the central heating source, approximated with a photosphere: its luminosity $L_{\star}$, its effective temperature $T_{\rm eff}$ and its mass $M_{\star}$. We estimate $L_{\star}$=13.6 $L_{\odot}$, $T_{\rm eff}$=4786 K and $M_{\star}$=3$M_{\odot}$ for Elias29 and $L_{\star}$=2.6 $L_{\odot}$, $T_{\rm eff}$=3980 K and $M_{\star}$=0.6$M_{\odot}$ for WL12, following the procedure discussed in Appendix \ref{AppendixC}. Then, a characterization of the disk structure is needed, i.e. the inner and the outer radii $R_{\rm in}$ and $R_{\rm out}$, the disk inclination angle $i$ (90$^{\circ}$ for an edge-on disk) and the dust surface density profile, that, for simplicity, is assumed to be a power-law:
\begin{equation}
\Sigma=\Sigma_0 \,\Biggl(\frac{R}{\text{1 AU}}\Biggr)^{-p},
\end{equation} 
where $\Sigma_0$ is the surface dust density at a radial distance of 1 AU from the central object and where we set $p=1$, since the quality of our data does not allow us to discriminate between different values of $p$. The value of $\Sigma_0$ was scaled to accomodate the total disk mass $M_{\rm disk}$. The disk inclination angle is assumed to be $i=60^{\circ}$, because the corresponding disk luminosity is equal to the one obtained by averaging over all possible inclination angles \citep[see e.g.][]{Butner94}.

The mm-SED is not sensitive to $R_{\rm in}$, which we then set equal to 0.25 AU. The measured fluxes at 1.1 and 3 mm are instead sensitive to the dust density, so that $R_{\rm out}$ and the mass of the dust in the disk $M_{\rm disk}$ can be constrained by the observations (assuming a dust opacity and a gas to dust mass ratio). We fix gas/dust to 100 by mass and we explore the effects of using different prescriptions for $\kappa_{\nu}$ depending on the dust size.

\subsection{Envelope models}
In order to model the envelopes of our sources we adopted the theoretical structure of a rotating and collapsing spheroid as derived by \cite{Ulrich}, as done by \cite{Crapsi} and \cite{Lommen}. This model assumes that matter crosses a spherical surface $r_0$  at a uniform rate and then it falls freely, conserving angular momentum. The surface at $r_0$ rotates uniformly.

The envelope density structure for a rotating and collapsing spheroid is given by:
\begin{equation}
\rho_{\rm env}(r,\theta)=\rho_0 \Biggl(\frac{R_{\rm rot}}{r}\Biggr)^{3/2}\Biggl(1+\frac{\cos\theta}{\cos\theta_0}\Biggr)^{-1/2}\Biggl(\frac{\cos\theta}{2\cos\theta_0}+\frac{R_{\rm rot}}{r}\cos^2\theta_0\Biggr)^{-1}\,,
\label{rho_env}
\end{equation}
where $R_{\rm rot}$ is the centrifugal radius  of the envelope, $\rho_0$  is the density in the equatorial plane at the centrifugal radius and the quantity $\theta_0$ is the solution of the parabolic motion of an infalling particle given by: $[r\,(\cos\theta_0 - \cos\theta)/(R_{rot}\cos\theta_0\sin^2\theta_0)]=1$. Equation \ref{rho_env}
represents a free-falling envelope on large scales and flattened on smaller. On
larger scales it therefore matches the assumptions in \cite{Jorgensen}.

The outer radius of the envelope is fixed to 50000 AU, the value of the centrifugal radius $R_{\rm rot}$ is 300 AU as found by \cite{Lommen} for Elias29, while $\rho_0$ is a free parameter used to adjust the envelope mass $M_{\rm env}$. In our envelope models we do not include outflow cavities, as done by \cite{Lommen}.

For each of the two sources, the temperature structure of the envelope was computed using the Monte Carlo Radiative Transfer code RADMC-3D \citep{RADMC3D}, implementing the envelope density structure of eq. (\ref{rho_env}). The source of heating at the center was the disk-star system, whose emission was previously computed using the two-layered disk model \citep{Dullemond} and the disk not included specifically in the 3D dust radiative transfer models used for the envelope structure. The source of heating has been implemented in a similar way as the one discussed in \cite{Butner94}.  

\subsubsection{Dust Opacity} 
The dust opacity depends on grain sizes, shape and composition \citep[e.g.][]{Miyake,Pollack,Draine}. Estimates of the level of grain growth from the mm-SED can be made by making assumptions on the chemical composition and shape of the dust grains.

During the early stages of planet formation dust grains are thought to either stick or fragment, depending on their relative velocity. Therefore it is important to consider a dust population characterized by a distribution of grains
with different sizes. We adopt a truncated power law $n(a) \propto a^{-q}$, between a minimum and a maximum size, respectively $a_{\rm min}$ and $a_{\rm max}$ \citep[see][]{Birnstiel11}. In this paper, we consider the grains having a spherical geometry, since the growth of small particles to larger ones (with sizes of $\sim$ 1 mm) typically leads to compact structures, that are closer to spheres \citep{Beckwith00}.
Fixed the chemical composition and the shape of the grains, the dust opacity law depends only on $q$, $a_{\rm min}$ and $a_{\rm max}$.

We set $a_{\rm min} \sim 0.1 \, \mu$m, because grains are supposed to grow from sub-micron ISM-like grains \citep{Mathis} and can be subject to collisional fragmentation, to a distribution in size of the grains population \citep{Birnstiel11}. Moreover $a_{\rm min}$ can be set equal to a fixed value, since the dependence of the millimeter dust opacity on it is very weak, as long as $a_{\rm min} \ll 0.1$ mm. We set $q=$3 and therefore $a_{\rm max}$ remains the only variable to determine. The dependence on $q$ is discussed in \cite{NattaTesti04}. The values of the dust opacity per unit dust mass $\kappa_{\nu}$ at 1.1 mm and 3 mm are 2.47 $\rm cm^{2} g^{-1}$ and 1.59 $\rm cm^{2} g^{-1}$ respectively.

\subsection{Model fitting}
Our aim is to find the set of parameters to simulate the sky brightness distribution that fits best the observations. The free parameters for the disk are its mass $M_{\rm disk}$, its outer radius $R_{\rm out}$ and $a^{\rm d}_{\rm max}$, that determines the disk dust opacity; for the envelope the free parameters are its mass $M_{\rm env}$ and its dust opacity, i.e. $a^{\rm env}_{\rm max}$.

Once  the simulated images are produced, they need to be compared with the interferometric observations.
To do that, we interpolate the model output images at the exact wavelengths of our observations.

It is not appropriate to compare the model images with the observed images, as the
interferometer acts as a spatial filter. Therefore it is necessary to simulate
interferometric observaitons of the model images, to account for the interferometer
primary beam attenuation and the filtering effect on the Fourier plane. To do this,
we firs multiply the model images with the primary beam attenuation pattern of the
interferometer (assuming a gaussian beam for both interferometers) and then we
perform the Fourier Transform of the attenuated model images and sample them at the
appropriate (\emph{u,v}) locations in order to obtain simulated visibilities.

Our procedure starts with fitting the disk emission: using the two-layered disk model \citep{Dullemond} we vary $M_{\rm disk}$, $R_{\rm out}$ and $a_{\rm amx}^{\rm d}$ to reproduce together $F_{\rm disk}^{\rm 1mm}$ and $F_{\rm disk}^{\rm 3mm}$. In particular the three parameters need to reproduce not only the integrated fluxes at the two frequencies, but also to take into account the interferometric data at the long baselines (i.e. since the disks appear not resolved on the \emph{u-v} plane, the outer radii need to be $R_{\rm out} < 200$ AU).

Once the three parameters are found, we implement the output flux of the disk model as source of heating at the center of the envelope. Using the Monte Carlo Radiative Transfer code RADMC-3D \citep{RADMC3D}, we vary $M_{\rm env}$ and $a_{\rm max}^{\rm env}$, in order to reproduce at the same time the single-dish and the interferometric fluxes both at 1.1 mm and at 3 mm.
  
In Fig. \ref{vis_mod} we present the best fit for the observed visibilities, both at 1.1 mm and 3 mm for Elias29 and WL12. Within the signal to noise ratio of the observations and the simplified models adopted, the simulated $(u,v)$ profiles reproduce adequately the observed ones.
The parameters of the models that provide a best match with the data are presented in section \ref{discussion} and in Table \ref{par_mod}. The best models have been selected as to provide a reasonably good match with the data, a full $\chi^2$ minimization analysis is beyond the scope of this simplified modeling. We will however discuss the model uncertaintes below. \\


\begin{figure*}
   \resizebox{\hsize}{!}
             {\includegraphics[width=1.\textwidth]{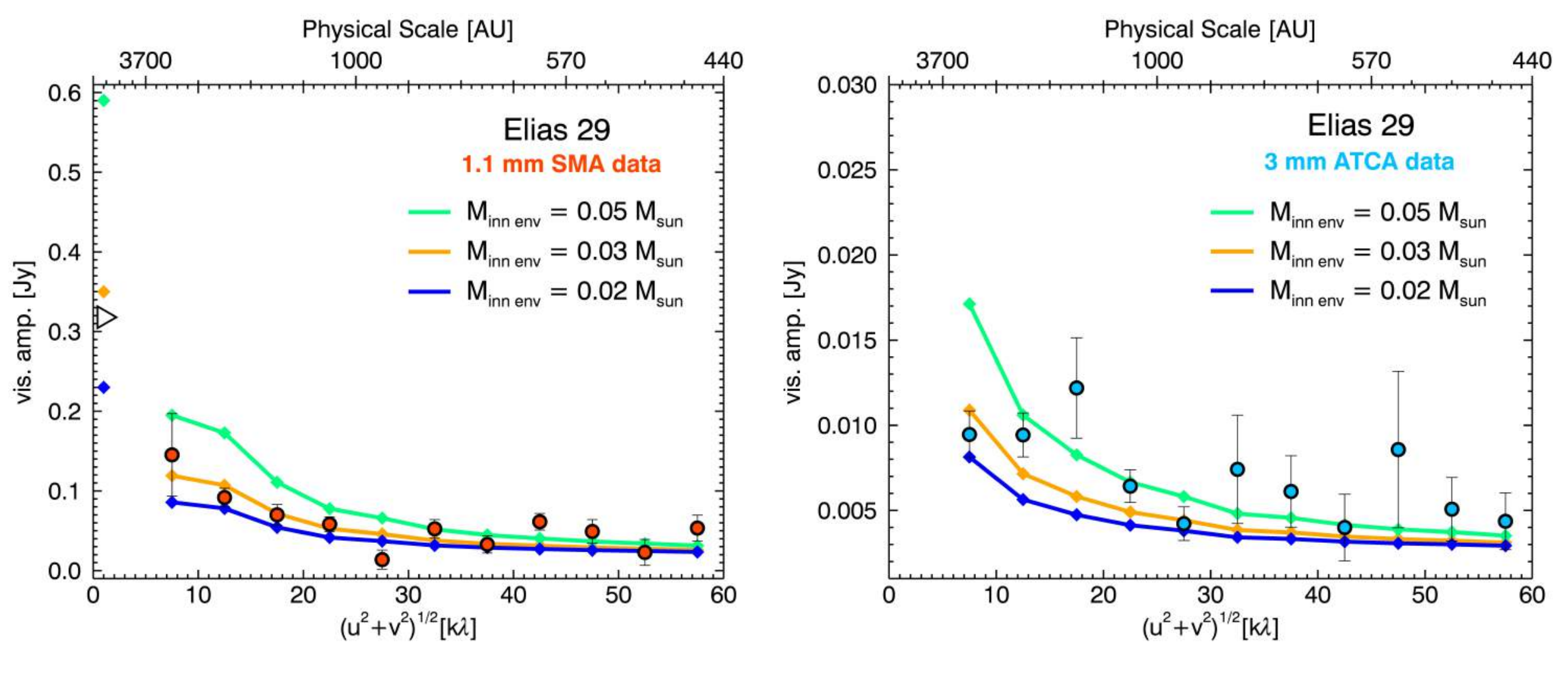}}
      \caption{Visibility amplitudes vs. \emph{u-v} distances for Elias29 at both 1.1 and 3 mm. The orange line is our best fit of the data, while the blue and green lines represent two models obtained decreasing and increasing the envelope mass respectively. Note that these two models also fail to reproduce the single dish flux at 1.1 mm \citep{Lommen,Jorgensen}, shown by the open triangle.}
         \label{mass_el29}
   \end{figure*}

\begin{figure*}
   \resizebox{\hsize}{!}
             {\includegraphics[width=1.\textwidth]{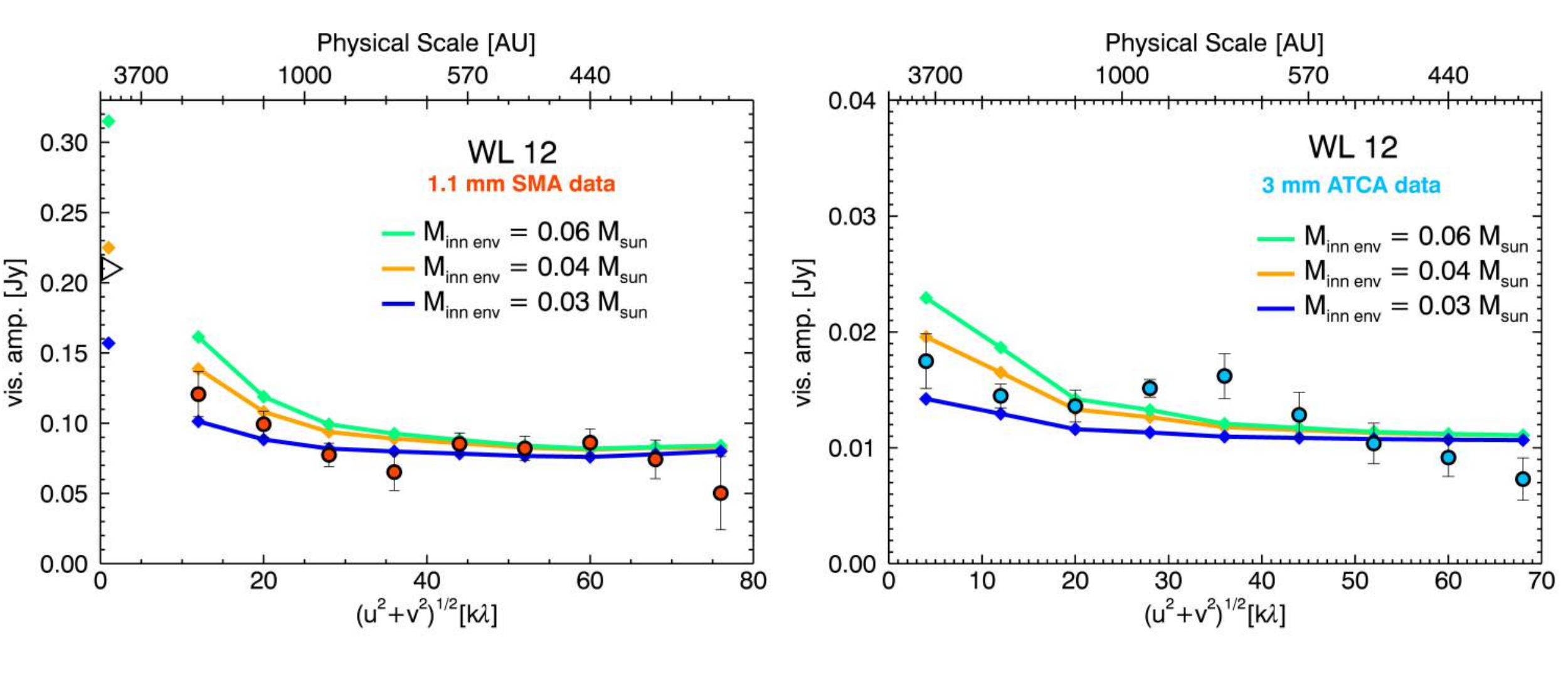}}
      \caption{Visibility amplitudes vs. \emph{u-v} distances for WL12 at both 1.1 and 3 mm. The orange line is our best fit of the data, while the blue and green lines represent two models obtained decreasing and increasing the envelope mass respectively. Note that these two models also fail to reproduce the single dish flux at 1.1 mm \citep{Lommen,Jorgensen}, shown by the open triangle.}
         \label{mass_wl12}
   \end{figure*}

\section{Discussion}
\label{discussion}
\subsection{Envelopes properties}
Since our interferometric data allow us to probe only the inner envelope regions, i.e. $R \lesssim 3000$ AU, with our simplified models we do not claim to give a detailed description of the whole envelope. For this reason we consider a large uniform envelope with a constant dust grain population, with the aim of reproducing the single dish fluxes and the behavior of the interferometric data on the \emph{u-v} plane. This approach is too simplistic to accurately model the outer envelope, infact one would need to compare more detailed models with observations at more compact baselines. Consequently our models to not reproduce the full observed SEDs of Elias29 and WL12: this kind of study is beyond the scope of our work.

The models that provide the best match to our observations were computed using the following parameters, reported also in Table \ref{par_mod}; a distribution of dust grains in the envelopes with maximum size $a_{\rm max}^{\rm env}$=1mm for both sources and inner envelope masses\footnote{We define the inner envelope mass as $M_{\rm inn\,env} \equiv M_{\rm env}(R < 3000 \,\text{AU})$, i.e. the mass enclosed in a sphere of radius $R=3000$ AU. } $M_{\rm inn\,env}\sim$ 0.03 $M_{\odot}$ for Elias29 and $M_{\rm inn\,env}\sim$ 0.04 $M_{\odot}$ for WL12. To estimate the uncertainties on the determination of the inner envelope masses, we show in Fig. \ref{mass_el29} and \ref{mass_wl12} the effects of changing this parameter. The orange line is our best match to the data, while the blue and the green lines represent the predictions by two models, where we decreased and increased the inner envelope mass in order to roughly reproduce within 1$\sigma$) the observed interferometric amplitudes. With this method we estimate an uncertainty in our derivation of the inner envelope mass of the order of 50\%\ (on top of the assumptions on the dust opacity coefficient and the gas to dust ratio mentioned above). The predicted single dish fluxes at 1.1 mm are represented by the coloured rhombi at zero \emph{u-v} distances in the left panels. The values obtained with the increased- and decreased-mass models highly overstimate and understimate the single dish fluxes, shown by the open triangles.

\begin{figure*}
   \resizebox{\hsize}{!}
             {\includegraphics[width=1.\textwidth]{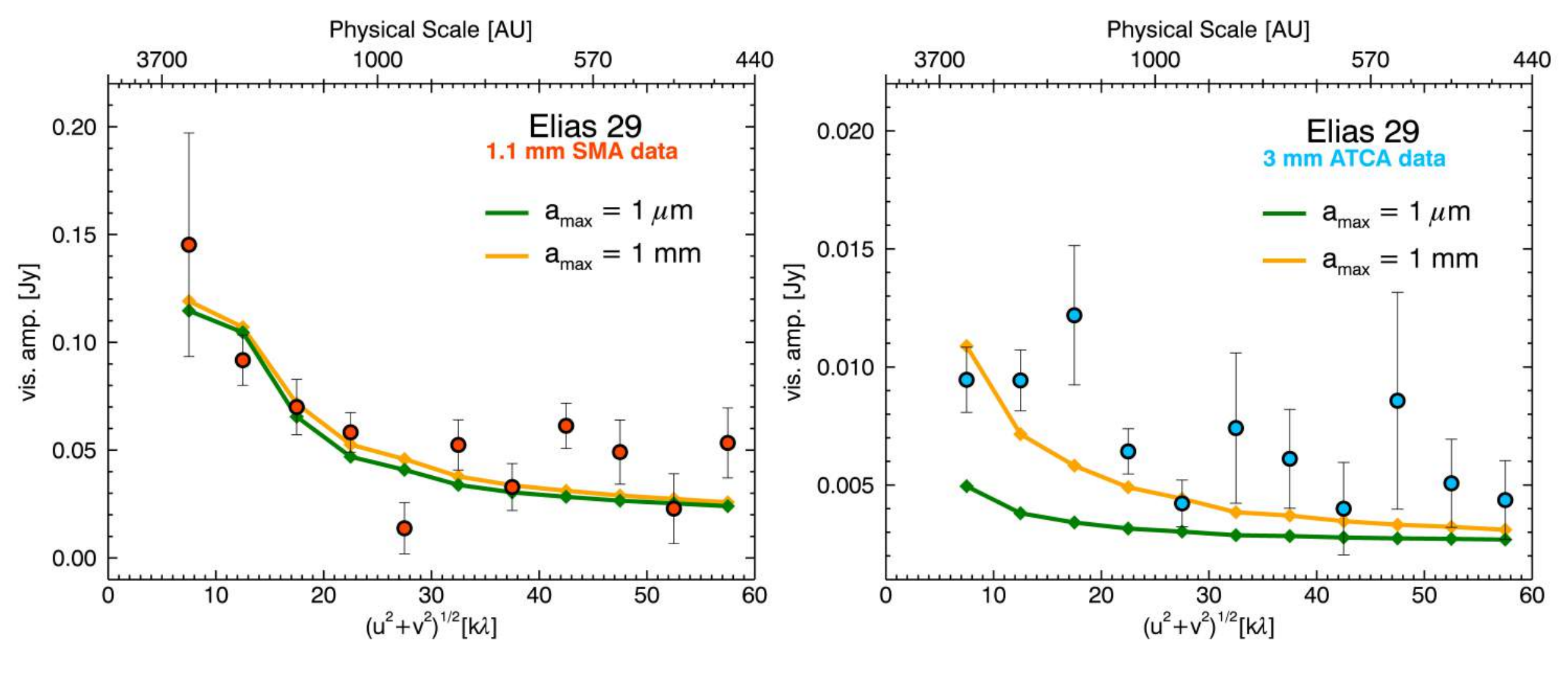}}
      \caption{Visibility amplitudes vs. \emph{u-v} distances for Elias29 at both 1.1 and 3 mm. The orange line is our best fit of the data, obtained with a 2 M$_{\odot}$ envelope, where grains have already been grown up to mm size. The green line represent a model obtained with a 1.1 M$_{\odot}$ envelope, with grains up to the micrometer size, in order to reproduce the 1.1 mm observations. It fails however to reproduce the 3 mm data. This shows that only the presence of already grown grains can give the dust opacity able to reproduce the dual-frequency observations.}
         \label{small_el29}
   \end{figure*}

We can constrain the level of grain growth in the envelopes within the framework of our physical but idealized collapsing rotating envelope models.
For example if we consider a dust grain size distribution in the envelope of Elias29 (see Fig. \ref{small_el29}), with a maximum size of 1 $\mu$m we can reproduce the 1.1 mm observations, but we very significantly understimate the 3 mm fluxes. 
The observations directly constrain the frequency dependence of the dust opacity coefficient $\beta$, and we find $\beta \sim$0.5--1. If we use our dust model to 
infer the maximum grain size from this value of $\beta$, we would imply the presence of dust grains grown to millimetre-size aggregates. As discussed by many authors, while the exact maximum size of grains depends on the assumption of the geometry and composition of grains, it is very difficult to explain the measured values of $\beta$ below 1 at millimetre wavelegnths without invoking a significant growth of the dust aggregates \cite[e.g.][]{NattaTesti04,Draine,Banzatti}.

The low values of $\beta$ in our sources are compatible with those found by \cite{Tobin13} in L1527 IRS. The fact that we find mm-sized grains in the envelope of Class I sources is also in agreement with the evidence of grown particles in the envelope of the Class 0 L1157-mm \citep{Chiang12}. These findings are consistent with the timeline proposed by the latest meteoritic studies for the growth of solids in our own Solar System: from chronological studies of solids found in meteorites, we know that the formation timescales of such mm-sized grains are similar to the median lifetimes of Class~0s and  Is \citep{Connelly12}. 
Grain growth in cores and protostellar envelopes has been modeled by \cite{Ormel09}, although current models cannot explain easily growth at the level we are inferring. 
The results of our analysis show that grain growth up to mm-size occurred in Elias29 and WL12 at scales between 2000 AU and 4000 AU. This corresponds, in our envelope models, to a range of number densities between $10^4$ to $10^5$ cm$^{-3}$. This is at least two orders of magnitude below the volume densities required by the models of \cite{Ormel09} for grains to grow to millimetre sizes in reasonable timescales. The models require $n > 10^6-10^7$ cm$^{-3}$ to form mm-sized grains on timescale of 1 Myr. 
This discrepancy needs to be analysed in more detail, both observationally and on the modeling side.

The possibility that grains can grow up to mm-size already in the collapsing envelopes of Class~0s and Is  is a result that may have important implications on the initial conditions for the models of dust evolution in protoplanetary disks. This would imply that the grain growth process within the disk starts with mm-sized particles, rather than with ISM-like grains as typically assumed by models of dust evolution in diks \citep{Birnstiel11}. However  this would probably not lead to very different conclusions on the dust evolution in the disk, since grain growth is expected to be a very fast process in the disk environment.

\begin{table}[ht]
\caption{Parameters derived from the modeling procedure.}
\centering
\begin{tabular}{lcc}
\hline \hline
\textbf{Source} & \textbf{Elias29} & \textbf{WL12} \\
\hline
$R_{\rm disk}$[AU] & $15 \textbf{-} 200$ & 30\\
$M_{\rm disk}[M_{\odot}]$ & $\gtrsim 0.01$ & $\gtrsim 0.3$\\
\textbf{$M_{\rm inn\,env}[M_{\odot}]$}  & $\sim$ 0.03 & $\sim$ 0.04 \\
$a_{\rm max}^{\rm env}$[mm] & 1 & 1\\
\hline
\label{par_mod}
\end{tabular}
\end{table}

\begin{figure}
   \resizebox{\hsize}{!}
		{\includegraphics[width=1.\textwidth]{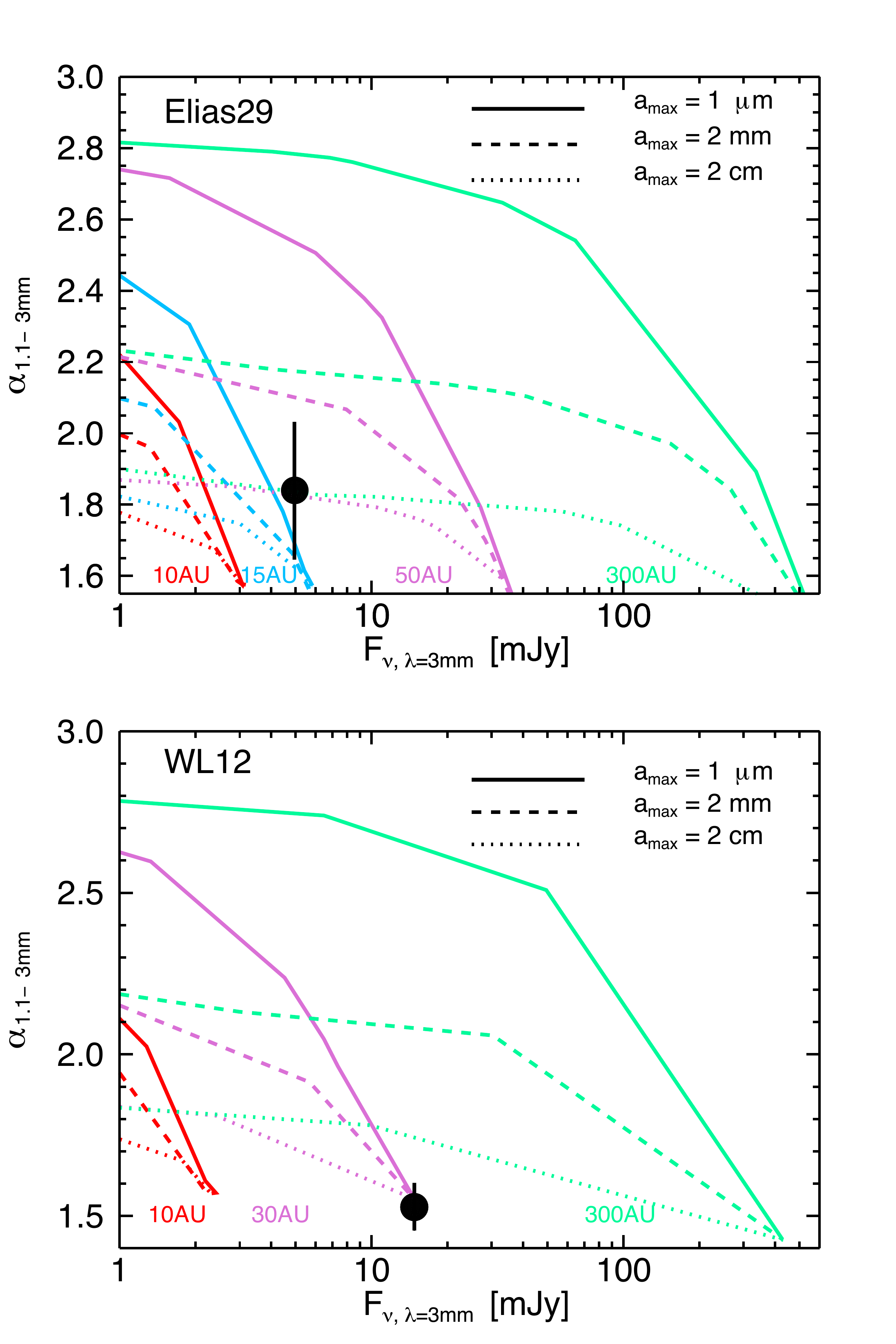}}
      \caption{Millimeter flux density vs. spectral index for disk models for Elias29 (top panel) and WL12 (bottom panel). The black dots show our ATCA data. Each line represents the prediction of disk models with the same disk outer radius and dust opacity spectral index $\beta$, but increasing disk mass from left to right.
For Elias29 the observed flux and spectral index are consistent with both a small (R$\sim$15 AU) optically thick disk, in which case we cannot constraint the dust properties, or a relatively large (R$\sim$50-200 AU) optically thin disk populated with cm-sized pebbles. For WL12 the match between the observed data and the models suggests a very small and optically thick disk with 30AU of radius. It is not possible to constrain whether the grains are in the $\mu$m , mm or cm size regime . }
         \label{alpha}
   \end{figure}

\begin{figure}
   \resizebox{\hsize}{!}
 {\includegraphics[width=1.\textwidth]{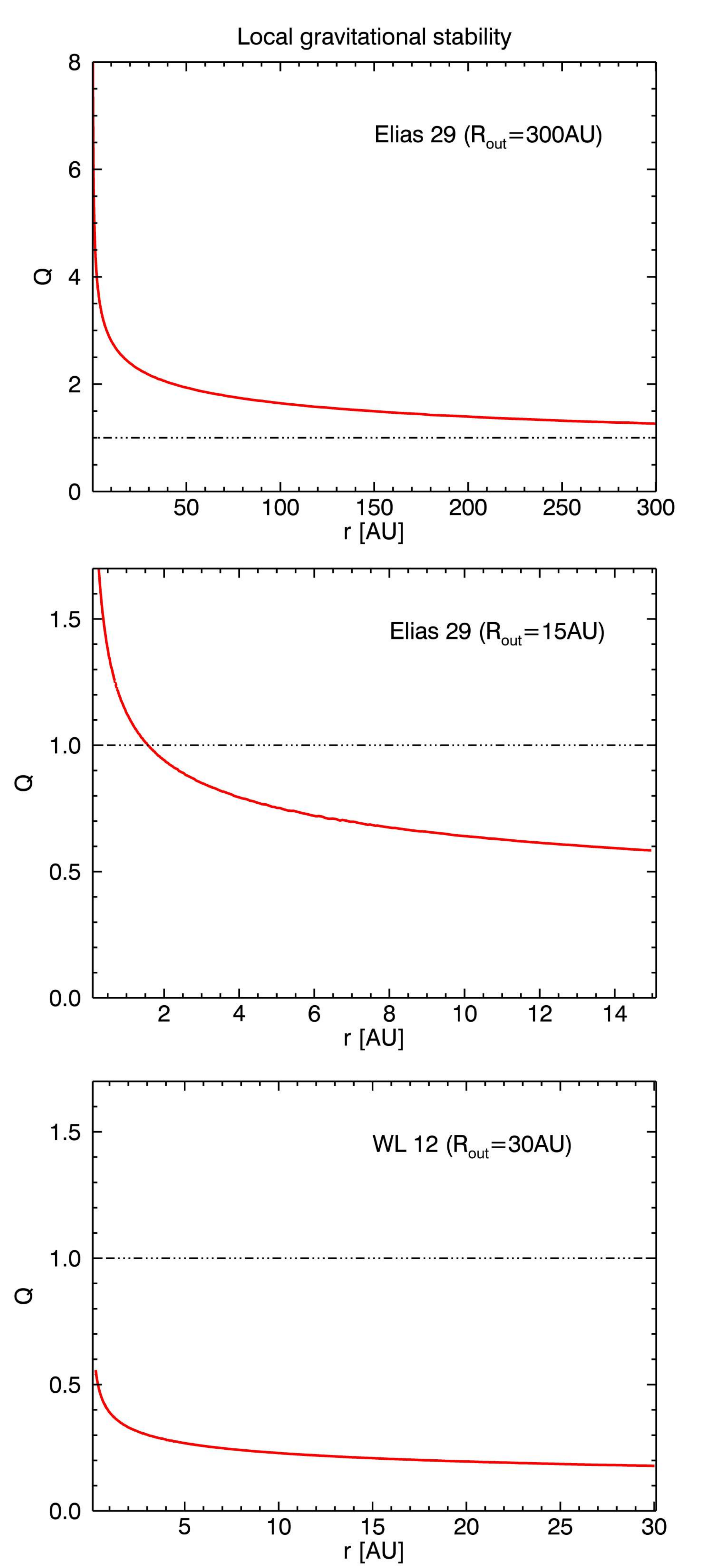}}
      \caption{Radial profile of the stability parameter $Q$. For Elias29 in the upper panel an extended disk model is assumed ($R_{out} = 300$ AU), while in the central panel a compact disk model is assumed ($R_{out} = 15$ AU). In the first case, in the outer regions of the disk $Q$ approaches unity, showing marginal stability. In the second case, the value of $Q$ drops below unity, showing the presence of strong gravitational instbility. In the bottom panel a compact disk model is assumed ($R_{out} = 30$ AU) for WL12. Also in this case $Q$ assumes very low values, showing that prominent gravitational instability is present.}
         \label{300AU}
   \end{figure}

\subsection{Disks properties}
For both sources the estimated value of $\alpha_{\rm disk}$ and consequently of $\beta_{\rm disk}$ are very low. To compare these results with the predictions of disk models we plotted the spectral index vs. the millimeter flux density for our data and for a set of families of disk models in Fig. \ref{alpha}, following the method developed in \cite{Testi01}. Different line colours represent the predictions of disk models with different disk outer radii. For each line the disk mass increases from left to right and accordingly also the mass surface density and the optical depth increase. For smaller disk masses the disk is optically thin and the flux increases when the mass increases. When the optical depth becomes $\gg 1$, the disk emission becomes less and less dependent on the disk mass and the spectral index assumes lower values. This explains the trend of the lines in the $\alpha$ vs $F$ diagram in Fig. \ref{alpha}. The predictions of models where we assumed different dust opacities, i.e. different grain sizes, are presented with different line styles. The overplotted black dots show our ATCA data for the two sources. 

The observed flux and spectral index of Elias29 are consistent with both a small ($R\sim$15 AU) optically thick disk, in which case we cannot constrain the dust properties, or a relatively large ($R\sim$50-200 AU) optically thin disk populated with large cm-sized pebbles. 
In the case of WL12, our data suggest a very small and optically thick disk with a 30 AU radius and lower limit for the mass of 0.3 M$_\odot$. Also in this case it is not possible to constrain whether the grains are in the $\mu$m , mm or cm size regime. We can however rule out the possibility of having a more extended disk with larger dust grains, since larger disks ($R_{\rm out}>$ 100 AU) would require solids as large as 1 meter to explain the low observed spectral index \citep{Ricci10}. This is however not supported by models of dust evolution in disks \citep{Birnstiel11}. The disks parameters derived from our analysis are summarized in Table \ref{par_mod}.

Another parameter that can be varied to match the observed spectral index shown in Fig. \ref{alpha} with a more extended disk, is the power-law index of the grains size distribution $q$, assumed to be $q=3$. Decreasing $q$, with no variation in the other free parameters, gives lower spectral index values. This decreasing trend stops however around $q=2.5$, as shown by \cite{Natta07}. If we decrease $q$ from 3 to the minimal value 2.5, we would obtain a spectral index just a factor of two lower \citep{Ricci10}. This difference is not big enough to require a substantialy larger disk. We therefore consider the hypothesis of an optically thick disk of 30 AU as the most plausible one.

Such compact disks however are not expected to form in the traditional scenario of disk formation if one neglects the effect of magnetic fields.
Due to the so-called angular momentum problem \citep{Spitzer78}, if the specific angular momentum is conserved during the collapse, the formation of large ($\gtrsim 100$ AU), rotationally supported disks is expected both in analytical models and in numerical simulations \citep{Tereby84,Yorke99}. On the other hand, the inclusion of  the magnetic fields can change significantly the picture. Due to magnetic braking, angular momentum is removed from the collapsing inner envelope and transferred at large radii, allowing the formation of only a small disk. In fact, some simulations run in the ideal MHD regime and for an initial alignment between the rotation axis of the core and the magnetic field lines found this process to be so efficient that it can even inhibit completely the formation of a disk \citep{Allen2003,Mellon2008,Hennebelle08}. The misaligment between the field lines and the rotation axis \citep{Hennebelle09,Joos12} can mitigate this effect. This misaligment was confirmed, at least in some case, from recent observations \citep{Hull13}.
Inclusion of the non-ideal MHD processes, namely ambipolar diffusion, Ohmic resistivity and Hall effect \citep{Dapp10,Krasno10,Krasno11,Li11,Dapp12}, also reduce the efficiency of magnetic braking, allowing in some cases the formation of small ($\sim10$ AU), rotationally supported disks. However, this is an open problem, and a clear consensus has still to emerge. 
If the size of Elias 29 and of WL12 were confirmed by higher angular resolution observations, this would put a strong constraint on models of disk formation.

Such compact and relatively massive disks are expected to be subject to gravitational instabilities. In order to estimate the importance of such processes, we compute the stability parameter $Q$ for both our sources, where $Q$ is defined as
\begin{equation}
Q=\frac{c_{\rm s}\kappa_{\rm ep}}{\pi G\Sigma}\approx \frac{c_{\rm s}\Omega}{\pi G\Sigma},
\end{equation}
where $\kappa_{\rm ep}\approx\Omega$ is the epicyclic frequency and $c_{\rm s}$ is the thermal sound speed. The radial profiles of $Q$ are shown in figure \ref{300AU}, for Elias29 for the two cases where an extended disk model is assumed ($R_{\rm out}\sim 300$ AU, upper panel) and where a compact disk model is assumed ($R_{\rm out}\sim 15$ AU, central panel), and for WL12 ($R_{\rm out}\sim 30$ AU, bottom panel). Even in the case where the extended disk model for Elias 29 is considered, we see that the disk approaches marginal stability ($Q\approx 1$) in the outer disk, implying that it can be weakly gravitationally unstable also in this case. For the two other cases considered, where the disk is more compact, the $Q$ parameter formally drops significantly below unity, indicating that some additional support against self-gravity needs to be present, for example in the form of turbulent motions generated by the instability. Indeed, gravitational instability is expected to provide turbulent motions at the sonic level \citep{LodatoRice04,Cossins09}, that for both sources would correspond to $v_{\rm turb}$ of the order of a few km/sec. Such turbulent phenomena induced by the disk self-gravity could thus represent an important source of angular momentum transport to bring Class I objects into the Class II phase. Additionally, gravitational instabilities are likely to result in strongly time-varying behaviour, such as episodic accretion \citep{Audard13} and/or fragmentation into bound objects, such as low-mass stellar companions or giant planets \citep{Rice05}. 


\section{Summary and Conclusion}
\label{Summary}

We present new 3 mm ATCA data of two Class I YSOs in the Ophiucus SFR: Elias29 and WL12. For our analysis we compared them with archival 1.1 mm SMA data \citep{Jorgensen}. In the \emph{u-v} plane the two sources present a similar behaviour: a costant non-zero emission at the long baselines, which shows the presence of an embedded unresolved disk and an increase of the fluxes at the short \emph{u-v} distances, related to the presence of an extended envelope. The main results from our analysis are the following.
\begin{itemize}
\item[-] \emph{Grain growth in the envelopes.} Our observations suggest that dust grains  start to aggregate up to mm sizes already in the envelope of the two Class I YSOs we have observed. This result is in agreement with the evidence of large grains in the envelope of the Class~0 L1157-mm \citep{Chiang12} and with chronological studies of solids found in meteorites \citep{Connelly12}. The fact that grains can grow up to mm-size already in the collapsing envelopes of Class~0s and Is is a result that may have important implications on the initial conditions for the models of dust evolution in protoplanetary disks. \\
\item[-] \emph{Optically thick disks.} From our modeling we conclude that the embedded disks in our Class Is are probably very compact, with outer radii down to tens of AU, at least for WL12.  The existence of such compact disks can suggest that magnetic fields are acting to remove the cloud angular momentum during collapse. The magnetic braking is more efficient during the first stages of collapse (Class 0, or early Class I YSOs) when the envelope is still massive. This may indicate that Elias29 and WL12 are young Class Is.  If the size of the two sources were confirmed by resolved observations, this would put a strong constraint on models of disk formation. Additionally, such compact disks are expected to be gravitationally unstable, which would result in large turbulent velocities, up to a few km/sec, and in the formation of a spiral structure in the disk, that might be potentially observed through ALMA \citep{Cossins10}.\\
\end{itemize}

Surely future higher resolution observations with the Atacama Large Millimeter/submillimeter Array (ALMA), which has the capability to resolve such compact disks, will be needed to accurately describe the structure of Class I YSOs. Moreover a larger sample of Class I sources, easily obtainable with ALMA's high sensitivity, would be essential to understand how general our results are.

\section*{Acknowledgements}
The authors wish to thank ATNF$\&$CSIRO staff for the support and the hospitality, F. Trotta and F. Testi for their help during the observing session. The authors also thank E. van Dishoeck and the referee, J. Jorgensen, for
insightful comments that significantly improved their work. This work was partly
supported by the ESO DGDF program and by the Italian Ministero dell’Istruzione,
Universita’ e Ricerca through the grant Progetti Premiali 2012 - iALMA.

\newpage

\begin{appendix}


\newpage
\section{Error Estimate of the Approximate Dust Opacity Spectral Index}
\label{AppendixB}

In the optically thin limit and the Rayleigh$-$Jeans regime, the dust opacity spectral index $\beta$ can be approximated using the flux density at two wavelengths. In this appendix, we discuss the error propagation from the observational uncertainty to
the deduced $\beta$ value as done by \cite{Chiang12}. Let $F_1$ and $F_2$ be the flux density at frequencies $\nu_1$ and $\nu_2$, $\beta$ can be expressed as in equation (\ref{beta_thin}):
\begin{equation}
\beta = \frac{\ln F_1 - \ln F_2}{\ln \nu_1 - \ln \nu_2} - 2\, .
\label{beta_thin2}
\end{equation}
Let us assume that the variables $F_1$ and $F_2$ are independent and that $\sigma_1$ and $\sigma_2$ are their standard deviations, propagating the errors we obtain:
\begin{equation}
\sigma_{\beta}^2=\Biggl|\frac{\partial \beta}{\partial F_1}\Biggr|^2\,\sigma_1^2 + \Biggl|\frac{\partial \beta}{\partial F_2}\Biggr|^2\,\sigma_2^2\, .
\label{sigma_beta}
\end{equation}
Calculating the partial derivative of equation (\ref{beta_thin2}) in both the variables $F_1$ and $F_2$ we obtain:
\begin{gather}
\frac{\partial \beta}{\partial F_1}=\frac{1}{(\ln\nu_1 -\ln\nu_2)\, F_1} ,\\
\frac{\partial \beta}{\partial F_2}=\frac{1}{(\ln\nu_1 -\ln\nu_2)\, F_2} .
\end{gather}
Combining these two results with eq. (\ref{sigma_beta}) we obtain the uncertainty on $\beta$:
\begin{equation}
\sigma_{\beta}^2 = \Biggl(\frac{1}{\ln\nu_1-\ln\nu_2}\Biggr)^2\Biggl(\frac{\sigma_1^2}{F_1^2}+\frac{\sigma_2^2}{F_2^2}\Biggr)\,.
\end{equation}


\section{Estimate of the Protostars Parameters}
\label{AppendixC}

In order to model the mm emission coming from the disks, the bolometric luminosity $L_{\text{bol}}$ and the effective temperature ${T_{\text{eff}}}$ of the central objects are needed. Once these two parameters are known, then one can obtain the black-body emissions related to the central protostars which are used as the input sources of energy for the computation of the disks dust temperature.

The bolometric luminosities were obtained with Spitzer photometry from the "Cores to Disk" legacy program \citep{Evans03} and are shown in Table \ref{par_birth}.
On the other hand, in order to obtain ${T_{\text{eff}}}$ some assumptions are needed. 

We assume our two Class I YSOs to lie along the stellar birthline, namely the locus in the H-R diagram along which young stars first appear as optically visible objects \citep{Stahler83}. We adopt the birthline constructed by \cite{Palla90} to obtain ${T_{\text{eff}}}$ given $L_{\text{bol}}$ \citep{Evans09} as inputs.

We expect Class I YSOs to have two important contributions from the central protostar: the photospheric stellar luminosity, $L_{\rm phot}$ (the one just considered) and the accretion luminosity, L$_{\rm acc}$. At this stage of evolution $L_{\rm acc}$ can be comparable with  $L_{\rm phot}$, while using this simple procedure we are assuming  $L_{\rm acc}\ll$ $L_{\rm phot}$, therefore negligible.\\
This could in principle make some differences in the results, in particular because $L_{\rm acc}$ tends to enhance the temperature and therefore to change the stellar spectrum. Since we do not have any kind of observation to constrain how much the accretion contributes to the total stellar luminosity, we assume to have only a photospheric contribution. This could lead to underestimate the temperature in the internal regions of the disk-envelope structure, and thus to overestimate the total dust mass. However we do not expect this effect to be crucial: \cite{Natta00} show that the average temperatures of disks around stars with very different $T_{\rm eff}$ change only by a factor of two. This means that our errors on the estimation of the dust mass given by the lack of informations on $L_{\rm acc}$ is only of a factor of two, which is comparable to the uncertainties we have on the dust opacities.\\
We can anyway check whether our results are reasonable or not, deducing the effective masses of the central objects and comparing it with the masses $M_s$ given by \cite{Jorgensen} and \cite{Lommen}, presented in Table \ref{par_lit}.\\
From  $L_{\text{bol}}$ and ${T_{\text{eff}}}$ we obtain $R_{\text{eff}}$ using the standard photosphere relation as follows:
\begin{equation}
R_{\text{eff}}= \Biggl( \frac{L_{\text{bol}}}{4\pi\sigma\, T_{\text{eff}}^4}\Biggr)^{1/2}.
\end{equation}
Then, to deduce $M_{\text{eff}}$,  we use the radius vs. mass relation obtained by \cite{Palla90} for a spherical protostar accreting at a rate of $10^{-5}\, \text{M}_{\odot}\text{yr}^{-1}$.\\
All the effective parameters obtained with the \emph{birthline method} are summarized in Table \ref{par_birth} together with the measured protostars masses already present in literature. The effective mass of Elias 29 is compatible with the measured ones. 

\end{appendix}

\end{document}